\documentclass[
reprint,amsmath,amssymb,aps,
]{revtex4-1}
\usepackage{graphicx}
\usepackage{dcolumn}
\usepackage{bm}
\usepackage{siunitx}

\begin{document}

\preprint{APS/123-QED}

\title{Thermal Fluctuations of Anisotropic Semiflexible Polymers}

\author{Sander L. Poelert}
\author{Harrie H. Weinans}
\author{Amir A. Zadpoor}%
 \email{a.a.zadpoor@tudelft.nl}
\affiliation{%
 Department of Biomechanical Engineering, Faculty of Mechanical, Maritime and Materials Engineering, Delft University of Technology, Mekelweg 2, Delft 2628 CD, The Netherlands
}%

\date{\today}

\begin{abstract}
Thermal fluctuations of microtubules (MTs) and other cytoskeletal filaments govern to a great extent the complex rheological properties of the cytoskeleton in eukaryotic cells. In recent years, much effort has been put into capturing the dynamics of these fluctuations by means of analytical and numerical models. These attempts have been very successful for, but also remain limited to, isotropic polymers. To correctly interpret experimental work on (strongly) anisotropic semiflexible polymers, there is a need for a numerical modelling tool that accurately captures the dynamics of polymers with anisotropic material properties. In the current study, we present a finite element (FE) framework for simulating the thermal dynamics of a single anisotropic semiflexible polymer. First, we demonstrate the accuracy of our framework by comparison of the simulated mean square displacement (MSD) of the end-to-end distance with analytical predictions based on the worm-like chain model. Then, we implement a transversely isotropic material model, characteristic for biopolymers such as MTs, and study the persistence length for various ratios between the longitudinal shear modulus, $G_{12}$, and corresponding Young's modulus, $E_{1}$. Finally, we put our findings in context by addressing a recent experimental work on grafted transversely isotropic MTs. In that research, a simplified static mechanical model was used to deduce a very high level of MT anisotropy to explain the observation that the persistence length of grafted MTs increases as contour length increases. We show, by means of our FE framework, that the anisotropic properties cannot account for the reported length-dependent persistence length.
\end{abstract}

\maketitle

\section{Introduction}
The characteristic thermal fluctuations of biopolymers that are submerged in a viscous solvent have been extensively studied for over a decade \cite{Brangwynne2007,Pampaloni2006,Gittes1993,Everaers1999}, mainly due to their important role in many biological processes \cite{Howard2001,Scholey2003}. In eukaryotic cells for example, the thermal undulations of biopolymers are essential for the functioning of the dynamic network that these biopolymers constitute \cite{franze2010biophysics, hatami2011}. This network is known as the cytoskeleton and mostly consists of the biopolymers F-actin, intermediate filaments, and microtubules, which are linked together by accessory proteins. The cytoskeleton gives the eukaryotic cell its mechanical properties, allows the cell to resist external stresses and plays a leading role in active force generation, cell mitosis and intracellular transport of vesicles and organelles \cite{Fletcher2010,hatami2011}. A good understanding of the mechanisms underlying these divergent roles of the cytoskeleton, requires a complete description of the mechanics and thermal dynamics of all of its structural components \cite{Bausch2006, Gardel2008}.

The thermal dynamics of a submerged filament is often discussed in terms of its persistence length, ${{\ell }_{p}}$. If the contour length, ${{L}_{c}}$, of the polymer is much greater than its persistence length, the dynamics of the polymer is dominated by thermal bending (flexible limit). Conversely, if the contour length is much smaller than the persistence length, the polymer dynamics will be dominated by elastic bending (stiff limit). Many biopolymers have a contour length close to their persistence length and are therefore called \emph{semiflexible}. Formally, the persistence length is the characteristic length scale of the arc length for which two tangent angles along the polymer's backbone $\theta (0)$ and $\theta (s)$ start to become uncorrelated \cite{Howard2001}. In two-dimensions, this translates to \cite{Gittes1993}:
\begin{equation} \label{eq:correlation_persistence}
\left\langle \cos (\theta (s)-\theta (0) ) \right\rangle =\exp ({-s}/{2{{\ell }_{p}})}\;.
\end{equation}By means of the equipartition theorem, the persistence length can be related to the bending rigidity of the polymer, $\kappa $, and the external thermal energy, ${{k}_{B}T}$, as \cite{Howard2001}
\begin{equation} \label{eq:persistencelength}
{{\ell }_{p}}= {\kappa }/{{{k}_{B}}T}\;.
\end{equation} 
For homogeneous isotropic filaments, $\kappa$ equals the product ($EI$) of the Young's modulus, $E$, and second moment of area, $I$. However, the effective bending stiffness of \emph{anisotropic} polymers is nontrivial and cannot be expressed in terms of individual material properties \cite{Gardel2008}.

\begin{figure}[!ht]
  \centering
    \includegraphics[width=\columnwidth]{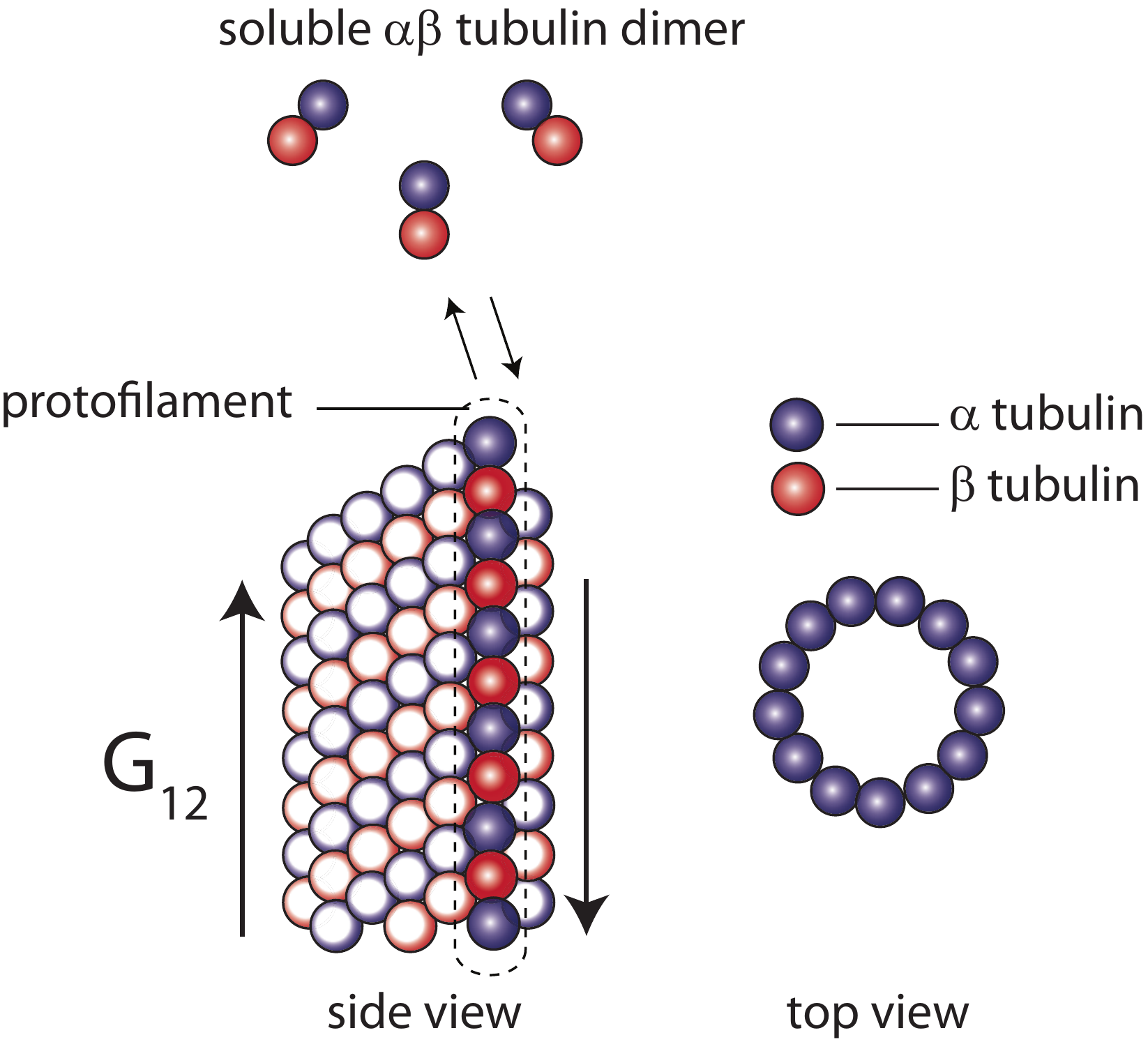}
  \caption[The microscopic architecture of a microtubule]%
  {A schematic image of the microscopic architecture of microtubules. Microtubules are straight, hollow filaments and consist out of, on average, 13 parallel aligned protofilaments.}
\label{fig:MTstructure}
\end{figure}

Microtubules (MTs) are among the cytoskeletal biopolymers that have attracted particular interest during the recent years, mainly due to their key role in cell division \cite{Jordan2004}. They exhibit an intrinsic anisotropic material behaviour \cite{Pampaloni2006,Gittes1993,Kis2002}. Underlying this anisotropic material behaviour is their complex molecular structure: MTs are hollow tubes with an outer diameter of \SI{25}{\nano\meter} that consist of parallel aligned protofilaments of $\alpha$- and $\beta$-tubulin heterodimers (Figure \ref{fig:MTstructure}). The longitudinal Young's modulus is determined by the strength of the head-tail $\alpha\beta$-$\alpha\beta$ tubulin bonds, whereas the shear modulus is determined by the much weaker inter-protofilament bonds \cite{pablo2003, schaap2004, sept2003}. All-atom computer simulations \cite{Sept2010} and \textit{in vitro} mechanical testing of MTs \cite{Kis2002} report anisotropic properties with one order difference between the longitudinal Young's modulus and corresponding shear modulus. 

Mainly due to the unfathomable relationship between anisotropic material properties of a polymer and its persistence length, it has been difficult to interpret recent experiments on grafted MTs submerged in a viscous fluid. In those experiments, a hidden dependency of the persistence length of grafted MTs on their contour length was observed \cite{Pampaloni2006,Taute2008}. Based on a simple Timoshenko beam model, that describes macroscopic elastic beams with shear contributions \cite{Gere1997}, and by assuming a static tip load and small displacements, anomalous MT material properties of a six orders lower longitudinal shear modulus than corresponding Young's modulus were found. Various numerical modelling studies indeed confirmed contour length-dependency of the persistence length, but these studies are likewise limited to static loads and small displacements \cite{An2010, Kasas2004, Li2006}. In this respect, there is a lack of an intuitive computational modelling tool that allows for the study of the thermal dynamics of anisotropic polymers under distributed thermal force application and finite displacements. 
  
Conventionally, polymers have been modelled numerically as a series of interconnected beads and rods or springs. This approach has its limitations, such as the lack of careful mathematical analysis, unavoidable artificial constraints and the need for (expensive) explicit time integration schemes \cite{Cyron2009}. Furthermore, bead models are limited to very slender and isotropic filaments, which makes these models less suitable for modelling the dynamics of relatively thick and anisotropic polymers, such as MTs. Driven by these limitations, a new method based on the finite element (FE) method was recently proposed \cite{Cyron2009}. In such an FE model, the microstructural complexity of a polymer is approximated by a continuum mechanical model, which is discretized into finite elements. The time-dependent partial differential equations (PDE's), resulting from the interaction between the polymer and its fluid environment, are solved over each element using implicit time integration. It can be proven mathematically that the solution of the FE method converges to the solution of the analytical PDE for infinitely small elements and time steps \cite{Cyron2009}. Additionally, FE modelling has already been used for many decades in other research fields, such as computational engineering. This has resulted in versatile user-friendly software packages, allowing straightforward implementation of advanced material models. Furthermore, the FE method has been shown to be up to thousand times faster than traditional bead-rod models based on explicit time integration schemes \cite{Cyron2009}. The sound mathematical formulation, easy implementation of complex material models and favourable computational costs makes the FE method an accurate and intuitive modelling tool for capturing the thermal dynamics of single semiflexible polymers.
 
In the current study, we develop a finite element framework, based on a commercial solver (Abaqus/standard, \textsc{simulia}), to model semiflexible polymers in thermal equilibrium with their viscous fluid environment. We first demonstrate the accuracy of this technique by comparing the simulated time evolution of the thermal fluctuations of freely floating and hinged isotropic filaments with analytical predictions based on the worm-like chain (WLC) model. We then change the boundary conditions to simulate a grafted MT and use the mean square of the transverse displacement of its tip as a measure for its persistence length. Again, we show good correspondence with the theoretical predictions based on the worm-like chain (WLC) model. Finally, we implement highly anisotropic material properties and study the persistence length of MT of various lengths. We relate our findings to recent experimental research \cite{Pampaloni2006}.

\section{Methodology}

\subsection*{Introduction}
In this section, we describe the numerical framework for capturing the thermal dynamics of single anisotropic semiflexible polymers. This numerical framework is built around a commercial FE solver (Abaqus/Standard, \textsc{simulia}) and is largely based on the framework proposed by Cyron and Wall \cite{Cyron2009}. The finite element approach is, as argued by Cyron and Wall in \cite{Cyron2009,Cyron2010}, an accurate, efficient and intuitive way of modelling the thermal dynamics of semiflexible polymers.  In the current section, we first explain the workflow of the framework centred around its three main stages (pre-processing, solving and post-processing). Then, to confirm our approach, we pay particular attention to the validation of the framework. Based on the time evolution of the mean square displacement (MSD) of the end-to-end distance of simulated isotropic MTs, we show good correspondence between simulation output and the analytical solution based on the WLC-model. We do this for various parameters and boundary conditions. Finally, we introduce an anisotropic material model and adapt the validated framework to address a recent experimental study on grafted anisotropic MTs \cite{Pampaloni2006}. 

\subsection{The finite element framework}
From here on, we coarse-grain the exact atomic architecture of an MT by approximating it as a hollow slender continuum structure, as is common practice in Brownian dynamics modelling \cite{Pampaloni2006,Howard2001}. Let us formulate a force equilibrium per unit length of internal and external forces to which the slender continuum is subjected \cite{Cyron2009}, 
\begin{equation} \label{eq:chp2force_eq}
{{\mathbf{f}}_{\operatorname{inert}}}(\mathbf{\ddot{u}})+{{\mathbf{f}}_{\operatorname{int}}}(\mathbf{u},\mathbf{\dot{u}},\mathbf{x})={{\mathbf{f}}_{\text{ext}}}(\mathbf{x}).
\end{equation}
In Eq.\eqref{eq:chp2force_eq}, the internal forces, ${{\mathbf{f}}_{\operatorname{int}}}$, include bending (elastic) forces, ${{\mathbf{f}}_{\operatorname{el}}}(\mathbf{u},\mathbf{x})$,  and damping forces, ${{\mathbf{f}}_{\operatorname{visc}}}(\mathbf{u}\mathbf{,\dot{u}}\mathbf{,x})$. Since we do not consider any deterministic external forces, the right hand side of Eq.\eqref{eq:chp2force_eq} only includes stochastic forces, ${{\mathbf{f}}_{\operatorname{stoch}}}(\mathbf{x})$. Furthermore, from dimensional analysis it can be shown that on the scale of a polymeric system the inertial forces are negligable compared to the internal friction, elastic and stochastic forces \cite{Doi1998}. Therefore, from here on we will not explicitly write the inertia force term. We thus rewrite Eq.\eqref{eq:chp2force_eq} as \cite{Cyron2009}
\begin{equation} \label{eq:chp2force_eq2}
{{\mathbf{f}}_{\operatorname{el}}}(\mathbf{u},\mathbf{x})+{{\mathbf{f}}_{\operatorname{visc}}}(\mathbf{u}\mathbf{,\dot{u}}\mathbf{,x})={{\mathbf{f}}_{\operatorname{stoch}}}(\mathbf{x}).	
\end{equation}
Eq.\eqref{eq:chp2force_eq2} is a nonlinear partial differential equation that cannot be solved analytically. We therefore resort to FE modelling for finding a solution for the displacement vector, $\mathbf{u}$. In order to efficiently perform the simulations, a modelling framework was built around a commercial FE solver (Abaqus/Standard, SIMULIA), based on a Fortran user-subroutine, MATLAB and Python-scripting (Table \ref{table:software}). 

\begin{table}[!ht]
\caption[Software used in the FE framework]{Overview of the software that was used in the finite element framework}
\centering
\small\begin{tabular}{c c c c} 
\hline\hline 
Software & Version & Manufacturer & Function \\ [0.5ex]
\hline
Python & 2.6.2 & Beopen.com & pre/post-processor \\
Abaqus/Standard & 6.10-1 & SIMULIA & solver \\
Fortran & 11 & Intel & solver \\
MATLAB & R2011a & Mathworks & post-processor \\ [1ex]
\hline
\end{tabular}
\label{table:software}
\end{table}

\begin{figure}[!ht]
  \centering
    \includegraphics[width=\columnwidth]{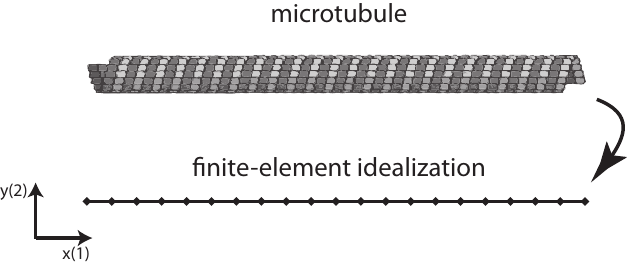}
  \caption[Biopolymer FE idealization]%
  {Initially the filament is aligned with the x-axis. Movement is confined to the x-y plane and the biopolymer (top) is discretized into a one-dimensional model of 20 elements (bottom). }
\label{fig:biopolymer}
\end{figure}

\subsubsection{Pre-processing}

Let us consider a polymer confined to movement in the x-y plane (2-D). At ${{t}_{\operatorname{sim}}}=0$, the backbone of the polymer is aligned with the x-axis, which can be considered the stress-free or reference configuration \cite{Poelert2011}. The contour length of the polymer, ${{L}_{c}}$, is a variable parameter and is specified by the user for each ensemble. The polymer is then discretised into 20 elements of length ${{l}_{e}}={{L}_{c}}/20$ (Figure \ref{fig:biopolymer}). Cyron and Wall (2009) showed already excellent results for discretization into 20 elements to capture the thermal dynamics of semiflexible polymers of similar properties as in the current study, and only minor improvements were reported a for more accurate discretization \cite{Cyron2009}. For discretization we used one-dimensional Timoshenko beam elements. This specific type of elements was chosen due to their ability to accurately describe shearing, even for relatively thick beams and low shear modulus \cite{Wei2008} and their favourable computational efficiency \cite{Poelert2011}. Additionally, several studies have shown that one-dimensional Timoshenko beam elements can accurately capture microtubule mechanics and are preferable over orthotropic shell models \cite{Shi2008,Gu2009}. The cross section of the filament is dependent on the type of polymer that is simulated and is presumed fixed throughout different ensembles \footnote{In many cases, we will be interested in an ensemble of simulations with equal parameters. From now on, we will call the run of a single simulation a \emph{realization} and the set of realizations with identical parameters an \emph{ensemble}.}.  Since we simulated MTs, which are of tubular shape, we assumed a hollow circular shaped cross-section, of which the dimensions are given in Table \ref{table:values}. We used one-dimensional continuum elements, thus the MT cross-section is only implicitly reflected by the second moment of area, $I$ \cite{Poelert2011}:
\begin{equation}
I=\pi \frac{r_{o}^{4}-r_{i}^{4}}{4},
\end{equation}
where $r_{o}$ is the outer radius and $r_{i}$ is the inner radius of the polymer.
First, for validation of the FE framework, we implemented an isotropic material model by setting the shear modulus to $G_{12}^{\operatorname{iso}}={{E}_{1}}/2(1+\nu )$. See Table \ref{table:values} for a listed overview of all the geometric and material properties that are used in this study.
\begin{table}[ht]
\caption{Values that were used for validation of the modelling framework}
\centering
\small\begin{tabular}{c c c c} 
\hline\hline 
Parameter &   & Value & Ref \\ [0.5ex]
\hline
Outer radius & $r_{o}$ & \SI{1.25e-2}{\micro\meter} & \cite{Howard2001} \\
Inner radius & $r_{i}$ & \SI{7.5e-3}{\micro\meter} & \cite{Howard2001} \\
Poisson's ratio & $\nu$ & $0.3$ & \cite{Sirenko1996} \\
Density & $\rho$ &  \SI{1.0e9}{\milli\gram\micro\per\meter\cubed} & \cite{Sirenko1996} \\ [1ex]
Young's modulus & $E$ & \SI{1.3e9}{\pico\newton\per\micro\meter\squared} & \cite{Wang2006} \\ [1ex]
Fluid viscocity & $\eta$ & \SI{1e-3}{\pico\newton\second\per\micro\meter\squared}   & \cite{Howard2001} \\ [1ex]
Thermal energy & $k_{B}T$ & \SI{4.045e-3}{\pico\newton\micro\meter}  & \cite{Cyron2010}  \\ [1ex]
\hline
\end{tabular}
\label{table:values}
\end{table}

Stochastic forces are applied on each individual node of the discretized polymer by means of a user-defined distributed load subroutine (DLOAD). At the beginning of each simulation time step $\Delta{t}_{\text{FE}}$ the DLOAD subroutine is called, upon which it outputs an array of random forces. These discrete forces are the equivalent of the noise correlations, fixed through the fluctuation-dissipation theorem  \cite{Cyron2009,Doi1998}, that determine the continuous stochastic force density $\textbf{f}_{\operatorname{stoch}}$. These random nodal forces are sampled from a Gaussian distribution, which is defined by its first and second moments,
\begin{equation} \label{eq:chp2:stoch_fluc}
\begin{array}{c}
\mu =0\\[10pt]
{{\sigma }^{2}}=2\frac{2{{k}_{B}}T\zeta }{\Delta {{t}_{\operatorname{stoch}}}L_{e}},
\end{array}
\end{equation} 
respectively, where we introduced the thermal energy of the solvent, ${{k}_{B}}T$, the drag-coefficient, $\zeta$, and the variable $\Delta {{t}_{\operatorname{stoch}}}$, which represents the `refresh rate' of the stochastic forces. The FE simulation step size, $\Delta {{t}_{\text{FE}}}$, has to be smaller than $\Delta{{t}_{\operatorname{stoch}}}$ in order to obtain proper FE convergence. In the current framework, we have consistently used a FE step size $\Delta {{t}_{\operatorname{FE}}}\le \left( 0.1\Delta {{t}_{\operatorname{stoch}}} \right)$.

Similar to earlier work \cite{Cyron2009}, we estimated the homogeneous isotropic drag coefficient, $\zeta $, according to the formula for a rigid cylinder in a homogeneous flow \cite{Doi1998} as, 
\begin{equation} \label{eq:chp2:zeta}
\zeta =4\pi \eta /\ln ({{{L}_{c}}}/{d)}\;,
\end{equation} where $\eta $ is the fluid viscosity and $d$ is the diameter of the filament \cite{Poelert2011}. The logarithmic term is a correction factor that compensates for the neglect of hydrodynamic interactions between distant segments of the polymer \cite{Doi1998,Chandran2009}. Although it has been demonstrated that contributions from internal friction cause a sharp increase in $\zeta $ for very short MTs, $\zeta $ may be presumed to obey Eq.\eqref{eq:chp2:zeta} in the range of ${{L}_{c}}$ considered in the current study \cite{Taute2008}. We modelled damping according to the Rayleigh damping model: 
\begin{equation} \label{eq:chp2:damping}
{{\mathbf{C}}_{(e)}}=\alpha {{\mathbf{M}}_{(e)}}+\beta {{\mathbf{K}}_{(e)}},
\end{equation}
where ${{\mathbf{C}}_{(e)}}$ is the elements damping matrix, ${{\mathbf{M}}_{(e)}}$ is the elements mass matrix and  ${{\mathbf{K}}_{(e)}}$ is the elements stiffness matrix. In our particular case of linear Timoshenko beam elements, we set $\alpha ={\zeta }/{\rho }\;A$, where $\rho$ is the density of the MT and $A$ is the cross-sectional area, and $\beta =0$.

\subsubsection{Solving}
Each realization was solved by the Abaqus/Standard FE solver on a Dell Precision T7500 workstation (Xeon X5680, two CPUs @ 3.33 GHz). The Abaqus/Standard FE solver is based on an efficient implicit time integration scheme, which makes it suitable for low-speed dynamic events, such as thermal undulations of polymers. The handling of the realizations (automatically starting the solver and collecting the data afterwards) is performed by a Python script.

\begin{figure}[!ht]
  \centering
    \includegraphics[width=\columnwidth]{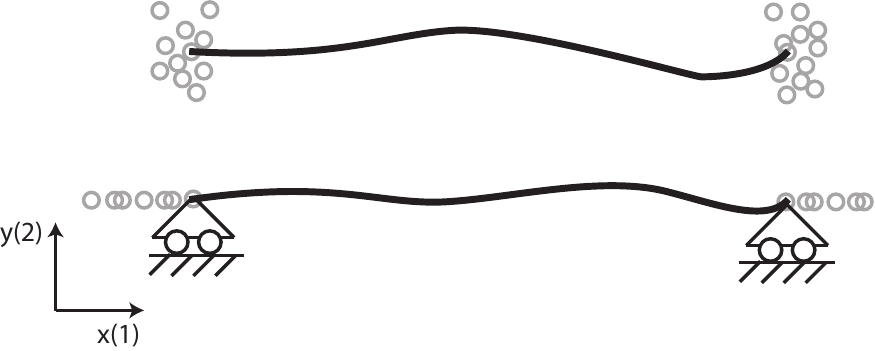}
  \caption[Boundary conditions for validation]%
  {Two sets of boundary conditions have been used for validation: a freely floating polymer (top) and a polymer with hinged ends (bottom). }
\label{fig:boundaryconditions}
\end{figure}

\begin{figure}[!ht]
  \centering
    \includegraphics[width=\columnwidth]{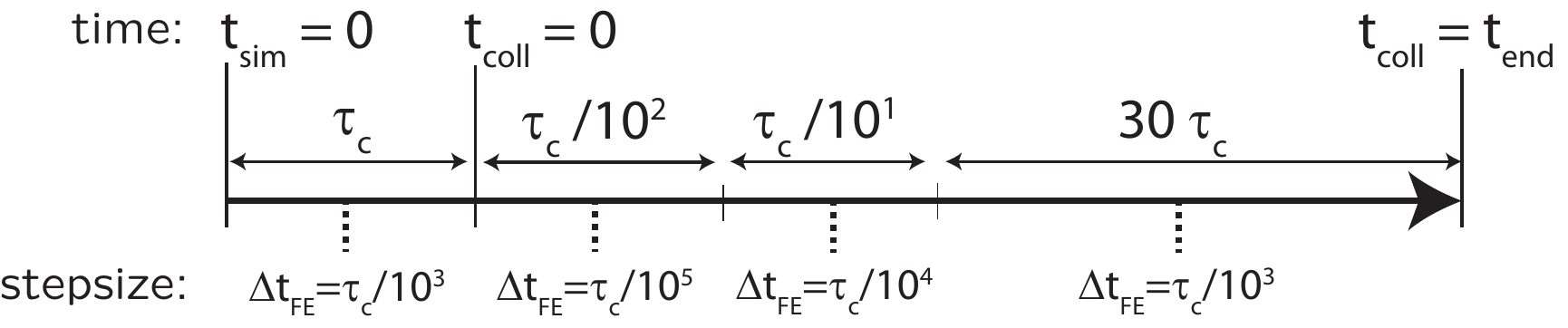}
  \caption[Timeline of simulation]%
  {A schematic depiction of the simulation timeline including the simulation time frame and collection time frame. The step size is adaptive: after relaxation is performed with a coarse step, the framework switches to a much smaller step size, which is then gradually increased.}
\label{fig:timeline}
\end{figure}

For solving the model, we introduced two different timeframes: the simulation time frame, ${{t}_{\operatorname{sim}}}$, and the data collection time frame ${{t}_{\operatorname{coll}}}$. At  ${{t}_{\operatorname{sim}}}=0$, the simulation is initiated and the initially straight contour of the filament starts to show bending fluctuations. To ensure that data collection is started after the polymer has reached equilibrium with its environment, we began data collection after the largest relaxation time of the polymer, ${{\tau }_{c}}$, elapsed, at which instant we set  $t_{\operatorname{coll}}$ (Figure \ref{fig:timeline}). This largest relaxation time of the polymer is determined by \cite{Gittes1993,Howard2001}
\begin{equation} \label{eq:chp2:longestrel}
{{\tau }_{c}}=\frac{\zeta }{{{\kappa }_{\operatorname{iso}}}}q{{_{1}^{*}}^{-4}}=\frac{\zeta }{EI}q{{_{1}^{*}}^{-4}},
\end{equation} where the wave vector of the first bending mode $q_{1}^{*}$ in Eq.\eqref{eq:chp2:longestrel} depends on the imposed boundary conditions. For example, for hinged boundary conditions $q_{1}^{*}={{{1.5\pi }/{L}\;}_{c}}$ \cite{Hallatschek2004} and for free boundary conditions $q_{1}^{*}={{{\pi}/{L}\;}_{c}}$ \cite{Howard2001} (Figure \ref{fig:boundaryconditions}). 

For various purposes, the dynamics need to be well described at very small timescales $\left( t<<{{\tau }_{c}} \right)$ as well as very large timescales $\left( t>>{{\tau }_{c}} \right)$ (e.g. for plotting the MSD of the end-to-end distance). In order to optimize the resolution for all timescales, variable time stepping is implemented: the time step that is used, depends on ${{\tau }_{c}}$ and the advancement of the realization. The procedure is as follows: first, the relaxation time is calculated based on the input parameters, Eq.\eqref{eq:chp2:longestrel}. A coarse time stepping resolution is used to let the polymer equilibrate with its fluid environment ($0<t<{{\tau }_{c}}$). Once the equilibrium has been obtained, the FE solver switches to the highest resolution and accurately captures the dynamics at the smallest timescale. The time resolution is gradually decreased, every decade of timescale, up to the largest time step that still allows for proper convergence of the realization (Figure \ref{fig:timeline}).

In the current FE framework, the computational cost of the simulation is, due to the variable time stepping, to a great extent independent of the relaxation time. Therefore, the computational effort is largely independent of the contour length of the polymer, the boundary conditions, the drag coefficient and the bending rigidity. However, for very large relaxation times or very low polymer rigidity the gain of variable time stepping is limited, due to a limit to the maximum step size that still allows for proper FE convergence. As an indication, a typical realization ($30\tau_{c}$ of simulated time) of an isotropic polymer lasted $\pm$ 45 minutes on the specified computer system. For high levels of anisotropy or very long relaxation times, the duration of a realization could take up to 4 hours.

The realizations within each ensemble are independent of each other (due to the ergodic nature of the studied phenomena). Therefore, simulations can be parallelized, reducing the overall simulation time. 

\subsubsection{Post-processing}
The solver outputs data files that contain a global time stamp of each step $\Delta {{t}_{\operatorname{FE}}}$ and the corresponding positions of the first and last nodes of the polymer.

\begin{figure}[!ht]
  \centering
    \includegraphics[width=\columnwidth]{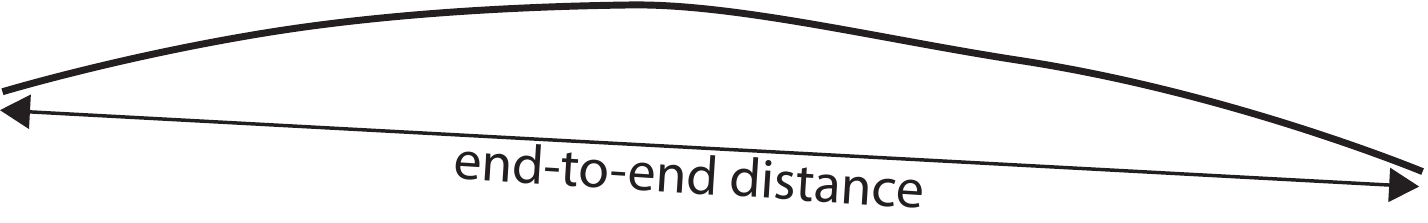}
  \caption[Linear shape functions]%
  {For the MSD of the end-to-end distance, that was used as observable in this framework, we collect the coordinates of the first and last node. For ${{t}_{\operatorname{coll}}}=0$ ${{\left\langle \delta {{R}^{2}}({{t}_{\operatorname{coll}}}) \right\rangle}}$ is set to zero and increases as time evolves.}
\label{fig:MSDendtoend}
\end{figure}

In order to quantify the time evolution of the polymer fluctuations, we define the mean square displacement (MSD) of the end-to-end distance as 
\begin{equation} \label{eq:chp2:MSD}
{{\left\langle \delta {{R}^{2}}({{t}_{\operatorname{coll}}}) \right\rangle }}\equiv {{\left\langle {{\left[ R({{t}_{\operatorname{coll}}})-R({{t}_{\operatorname{coll}}}=0) \right]}^{2}} \right\rangle }},
\end{equation} 
where $R\left( {{t}_{\operatorname{coll}}} \right)$ is the end-to-end distance or projected length of the filament (Figure \ref{fig:MSDendtoend}). As can be seen from Eq.\eqref{eq:chp2:MSD}, the MSD of the end-to-end distance is zero if ${{t}_{\operatorname{coll}}}=0$ and gradually increases with time. The end-to-end distance is determined by subtracting the position vector of the last node and the position vector of the first node of the polymer. This procedure is performed for all time steps $\Delta {{t}_{\operatorname{FE}}}$. The end-to-end distance at the  of the collection $R({{t}_{\operatorname{coll}}}=0)$ is subtracted from the total end-to-end distance and the resulting values are squared to obtain the displacement of the end-to-end distance $\delta {{R}^{2}}({{t}_{\operatorname{coll}}})$. We then average over all $N_{R}$ realizations in the ensemble to obtain the MSD of the end-to-end distance ${{\left\langle \delta {{R}^{2}}({{t}_{\operatorname{coll}}}) \right\rangle}}$.
\subsection{Validation results}
For validation purposes, two sets of boundary conditions have been considered: a freely floating polymer and a polymer with hinged ends (Figure \ref{fig:boundaryconditions}). In case of hinged-hinged boundary conditions, the end-nodes of the polymer are allowed to move along the x-axis, but are constrained to move along the y-direction.

According to the theory of the Brownian dynamics of polymers \cite{LeGoff2002,Hallatschek2004,Granek1997} two regimes can be identified in the time evolution of the mean square end-to-end distance: for very small times $t<<{{\tau }_{c}}$ \footnote{For the sake of clarity, from here on we omit the timeframe specifier subscript \textit{coll}, thus from here on $t\equiv t_{coll}$}, ${{\left\langle \delta {{R}^{2}}({{t}}) \right\rangle}}$ increases, obeying a power law that scales with ${{t}^{{3}/{4}\;}}$, whereas for $t>>{{\tau }_{c}}$, ${{\left\langle \delta {{R}^{2}}({{t}}) \right\rangle}}$ reaches a universal equilibrium value of ${{\left\langle \delta {{R}^{2}}(t) \right\rangle }_{\operatorname{eq}}}={{L}_{c}}^{4}/90{{\ell }_{p}^2}$ \cite{MacKintosh1995}. We rescaled the simulated ${{\left\langle \delta {{R}^{2}}({{t}}) \right\rangle}}$ and analytical solution by defining $F(t)=\left\langle \delta {{R}^{2}}(t) \right\rangle \left( 90{{\ell }_{p}^2}/{{L}_{c}}^{4} \right)$ and we rescaled time by defining $\tilde{t}=t/{{\tau }_{c}}$. For both boundary conditions (free and hinged) and contour lengths (\SI{10}{\micro\meter} and \SI{20}{\micro\meter}) a good agreement with the theoretical predictions of the WLC model \cite{Hallatschek2004,Granek1997} is observed for over 5 decades of timescale. The jagged shape of the simulated curves is due to the random nature of the observed phenomenon and is expected to die out completely for a higher number of realizations, as shown in a similar research \cite{Cyron2009}. It is not the goal of this particular study to show convergence of the finite element towards the analytical solution for a large number of realizations (this has been done before in \cite{Cyron2009} for ${{N}_{R}}=4000$). For the purpose of the current investigation, namely validation of the framework, ${{N}_{R}}=\pm 50$ realizations per ensemble was considered sufficient to show a good correspondence with the WLC theory.
\begin{figure}[!ht]
  \centering
\includegraphics[trim = 0mm 0mm 6mm 0mm, clip, width=\columnwidth]{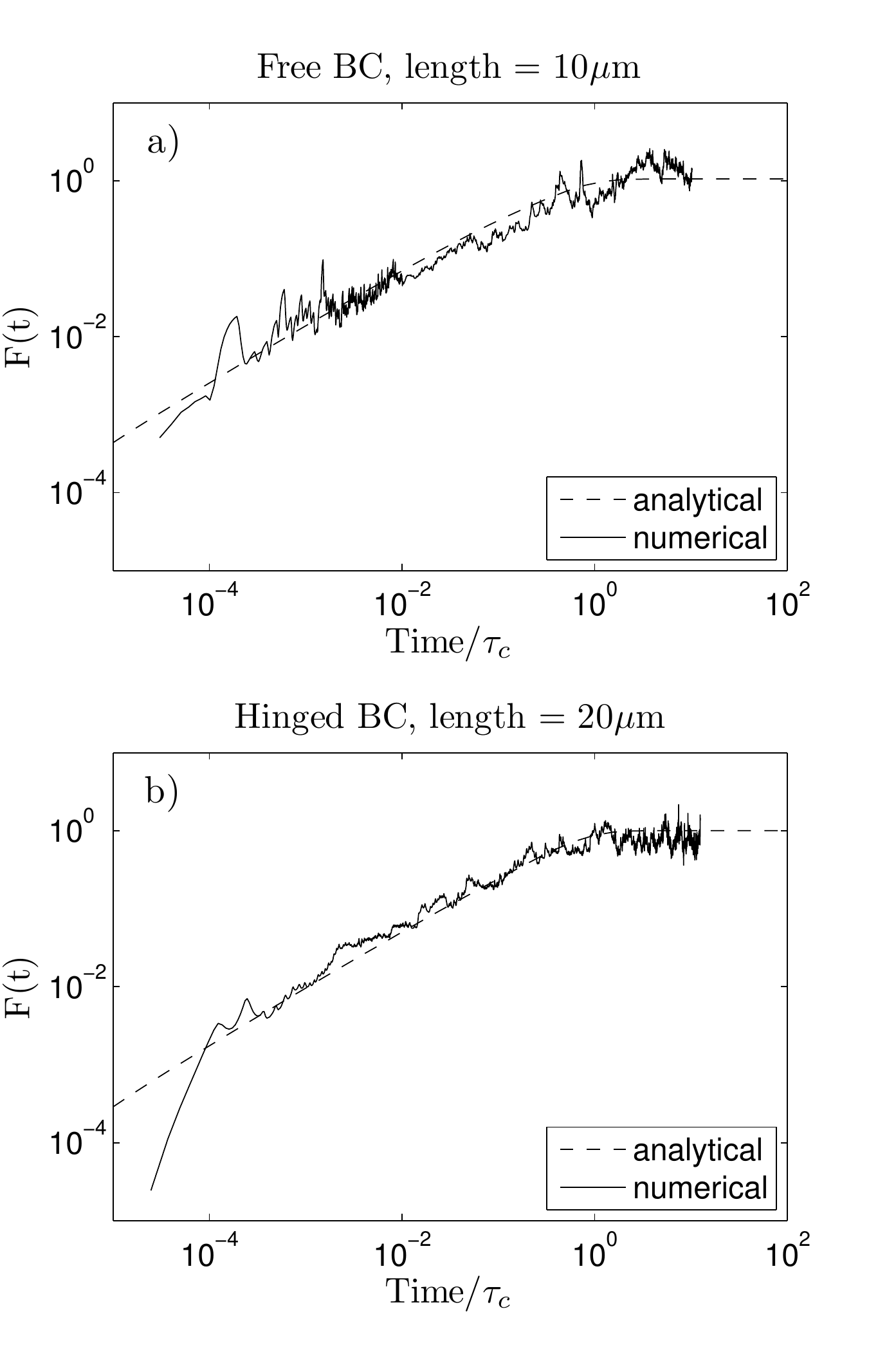}
  \caption[Persistence length versus contour length]%
  {Rescaled MSD of the end-to-end distance, $F(t)={\left\langle \delta {{R}^{2}}({{t}_{\operatorname{coll}}}) \right\rangle }\left( 90{{\ell }_{p}^2}/{{L}_{c}}^{4} \right)$, as a function of rescaled time $t/{{\tau }_{c}}$. The plots show the $F(t)$ for various combinations of boundary conditions and contour lengths [(a) BC: free, $L_{c}=\SI{10}{\micro\meter}$, (b) BC: hinged, $L_{c}=\SI{20}{\micro\meter}$]. }
\label{fig:chp2:validation_MSD}
\end{figure}
\subsection{Anisotropic polymers}
\begin{figure}[!ht]
  \centering
    \includegraphics[width=\columnwidth]{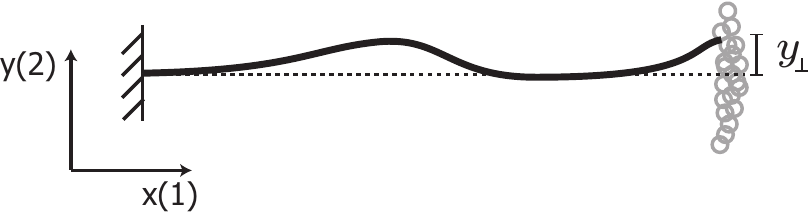}
  \caption[Clamped microtubule]%
  {Schematic drawing of the studied microtubule: the left-end is constrained in all degrees of freedom, whereas the right-end remains unconstrained. The mean square of the transverse displacement of the tip, $y_{\bot}$, is used as a measure for the persistence length.}
\label{fig:graftedMT}
\end{figure}

For the validation of the framework we presumed isotropic material properties. However, due to their atomic architecture (Figure \ref{fig:MTstructure}), MTs are known to be anisotropic: the longitudinal Young's modulus is determined by the strength of the head-tail $\alpha \beta -\alpha \beta$ tubulin bonds, whereas the shear modulus is determined by the much weaker inter-protofilament bonds \cite{pablo2003, schaap2004, sept2003}. Furthermore, in the plane of the transverse cross-section, MTs give an isotropic response to stress. From here on, we introduce a transversely isotropic material model, which accounts for the anisotropic material behaviour along the backbone of the MTs and isotropic properties along their cross-section. This approach is similar to other studies on MTs \cite{Pampaloni2006,Shi2008}.  According to linear elasticity theory, the elastic compliance matrix of a transversely isotropic material is defined by the Young's moduli in the plane of isotropy, ${{E}_{2}}={{E}_{3}}={{E}_{p}}$, the transverse Young's modulus ${{E}_{1}}={{E}_{t}}$ , the Poisson ratio's ${{\nu }_{p}}$, ${{\nu }_{pt}}$ , ${{\nu }_{tp}}$ and the in-plane and transverse shear moduli, ${{G}_{13}}={{G}_{23}}={{G}_{p}}$ and ${{G}_{12}}={{G}_{t}}$, respectively. Therefore, the 3-D stress-strain law reduces to:
\begin{equation} \label{eq:chp2:stressstrain}
\left( \begin{matrix}
   {{\varepsilon }_{11}}  \\
   {{\varepsilon }_{22}}  \\
   {{\varepsilon }_{33}}  \\
\end{matrix} \right)=\left( \begin{matrix}
   1/{{E}_{t}} & -{{\nu }_{tp}}/{{E}_{t}} & -{{\nu }_{tp}}/{{E}_{t}}  \\
   -{{\nu }_{pt}}/{{E}_{p}} & 1/{{E}_{p}} & -{{\nu }_{p}}/{{E}_{p}}  \\
   -{{\nu }_{pt}}/{{E}_{p}} & -{{\nu }_{p}}/{{E}_{t}} & 1/{{E}_{p}}  \\
\end{matrix} \right)\left( \begin{matrix}
   {{\sigma }_{11}}  \\
   {{\sigma }_{22}}  \\
   {{\sigma }_{33}}  \\
\end{matrix} \right)
\end{equation}	
\begin{equation}
\left( \begin{matrix}
   {{\gamma }_{12}}  \\
   {{\gamma }_{13}}  \\
   {{\gamma }_{23}}  \\
\end{matrix} \right)=\left( \begin{matrix}
   1/{{G}_{t}} & 0 & 0  \\
   0 & 1/{{G}_{t}} & 0  \\
   0 & 0 & 1/{{G}_{p}}  \\
\end{matrix} \right)\left( \begin{matrix}
   {{\sigma }_{12}}  \\
   {{\sigma }_{13}}  \\
   {{\sigma }_{23}}  \\
\end{matrix} \right)
\end{equation}	
Because of the 2-D confinement of the simulation, deflection of the filament is only determined by the longitudinal elastic modulus, ${{E}_{1}}$, and longitudinal shear modulus, ${{G}_{12}}$. Therefore, we introduced the transverse isotropy through the parameter, $\chi $, which sets the order of the ratio between the longitudinal shear modulus and corresponding Young's modulus, 
\begin{equation} \label{eq:chi}
\frac{{{E}_{1}}}{{{G}_{12}}}={{10}^{\chi }}.
\end{equation}
To conform our research as much as possible to the experimental work that has been done on grafted microtubules [2], we clamped the left-end of the MT by applying constraints in all three degrees of freedom (2 lateral, 1 rotational), whereas we left the right-end unconstrained (Figure \ref{fig:graftedMT}). This new set of boundary conditions changes the longest wave vector, which is now given by $q_{1}^{*}=1.875/{{L}_{c}}$ \cite{Wiggins1998}. Substitution of this wave vector in Eq.\eqref{eq:chp2:longestrel} gives us for an isotropic grafted polymer a relaxation time of
\begin{equation} \label{eq:chp2:relax_grafted}
\tau _{c}^{\operatorname{iso}}=\frac{\zeta }{EI}{{\left( \frac{{{L}_{c}}}{1.875} \right)}^{4}}.
\end{equation}	
For anisotropic polymers with a low shear modulus, it is expected that the effective bending rigidity decreases and thus the relaxation time increases. To estimate the effective relaxation time of an anisotropic polymer, $\tau _{c}^\text{eff}$, we studied the MSD of the end-to-end distance,${\left\langle \delta {{R}^{2}}({{t}_{\operatorname{coll}}}) \right\rangle }$, for each unique set of parameters. Based on the onset of the ${\left\langle \delta {{R}^{2}}({{t}_{\operatorname{coll}}}) \right\rangle }$ plateau we estimated $\tau _{c}^\text{eff}$. For subsequent simulations, data was only collected after this effective relaxation time plus an additional safety margin elapsed. We then considered for each ensemble the position of the grafted MT tip in the x-y plane which we sampled at an interval of $\tau _{c}^\text{eff}/{{10}^{3}}$. This allowed for accurate measurement of the distribution function $P(x,y)$ and transverse displacement ${{y}_{\bot }}$of the tip position (Figure \ref{fig:graftedMT}). Then, by integrating the probability distribution function over the y-axis, we found the asymmetric probability density function along the x-axis $P(x)$.

For semiflexible polymers in the stiff limit (${{L}_{c}}<{{\ell }_{p}}$), this asymmetric distribution is peaked toward full stretching and has a typical width of \cite{Lattanzi2004}:
\begin{equation} \label{eq:chp2:parallel}
{{L}_{\parallel }}={{L}^{2}}/{{\ell }_{p}}.
\end{equation}
Similarly, by integration along the x-axis, we obtained the distribution function $P(y)$, which is a Gaussian distribution centred at $y=0$, the variance of which is given by the mean square transverse displacement $\left\langle y_{\bot }^{2} \right\rangle$ \cite{Lattanzi2004}. This mean square transverse displacement $\left\langle y_{\bot }^{2} \right\rangle$ can be directly related to the persistence length of the polymer, according to \cite{Howard2001,Lattanzi2004,Keller2008}:
\begin{equation} \label{eq:chp2:lattanzi}
{{\ell }_{p}}=\frac{L_{c}^{3}}{3\left\langle y_{\bot }^{2} \right\rangle }.
\end{equation}
For each realization that is added to the ensemble, the mean square transverse displacement, and thus the persistence length, changes. The ensemble was considered to contain enough realizations for a reliable estimate of the persistence length, if for each consecutive addition of the last five realization to the ensemble the mean square transverse displacement did not change by more than 3\%.
\section{Results}
\begin{figure}[!ht]
  \centering
  \includegraphics[trim = 6mm 0mm 12mm 0mm, clip, width=\columnwidth]{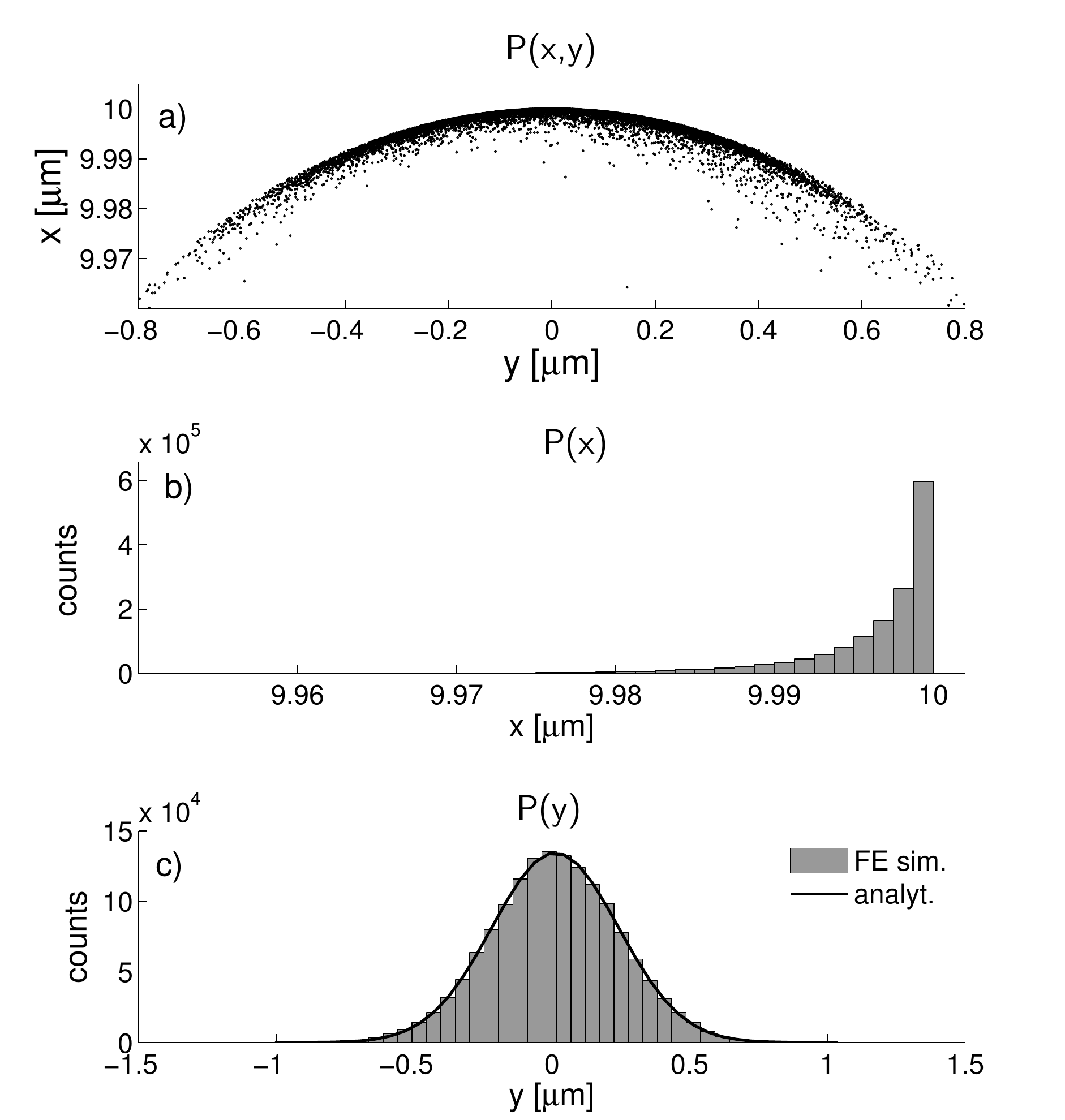}
  \caption[Point cloud and distribution functions of a grafted MT]%
  {a) Point cloud of sampled tip position, based on \SI{1.8e6} samples. b and c) After integration along the y and x-axis we find the two characteristic distributions $P(x)$ and $P(y)$, respectively. c) The solid line shows the analytical prediction according to the worm-like chain model. }
\label{fig:gaussian}
\end{figure}

First, to affirm our approach, we sampled the tip of grafted isotropic MTs of various lengths and calculated the simulated persistence length using Eq.\eqref{eq:chp2:lattanzi}. We found a good agreement with the theoretical persistence length (${{\ell }_{p}}=EI/{{k}_{B}}T= \SI{6.3}{\milli\meter})$. As an illustration, Figure \ref{fig:gaussian} shows the tip position of an isotropic MT of contour length ${{L}_{c}}=\SI{10}{\micro\meter}$ in the x-y plane, based on $1.8\times {{10}^{6}}$ samples. By integration along the y-axis, we find the characteristic non-Gaussian and asymmetric distribution function $P(x)$ with a width of ${{L}_{\parallel }}=\SI{0.015}{\micro\meter}$, as predicted by Eq.\eqref{eq:chp2:parallel} according to the WLC theory [31]. Similarly, integration along the x-axis results in the Gaussian distribution $P(y)$, for which a good agreement is observed with the WLC theory (Figure \ref{fig:gaussian}). Indeed, we find a mean square transverse displacement of $\left\langle y_{\bot }^{2}\right\rangle=\SI{0.0550}{\micro\meter}$, which is within 4\% accuracy of the theoretical value of $\left\langle y_{\bot }^{2}\right\rangle=0.0529$.

\begin{figure}[!ht]
  \centering
    \includegraphics[trim = 25mm 0mm 10mm 0mm, clip, width=\columnwidth]{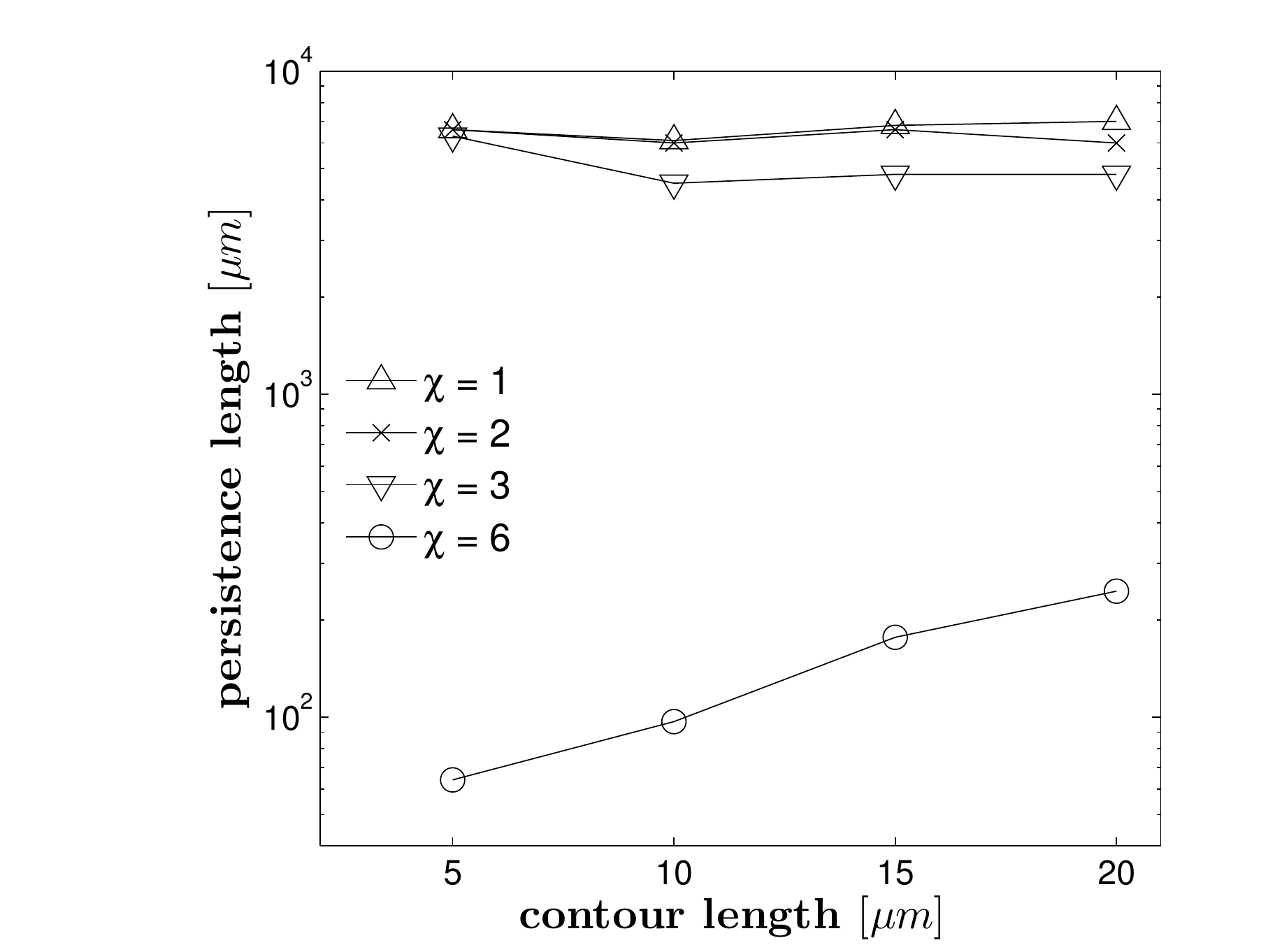}
  \caption[Persistence length versus contour length]%
  {Persistence length versus contour length for various levels of anisotropy. The parameter $\chi $ refers to the order of difference between longitudinal shear modulus and Young's modulus, see Eq.\eqref{eq:chi}.}
\label{fig:chp2:results}
\end{figure}

\begin{figure}[!ht]
  \centering
   \includegraphics[trim = 30mm 0mm 11mm 0mm, clip, width=1\columnwidth]{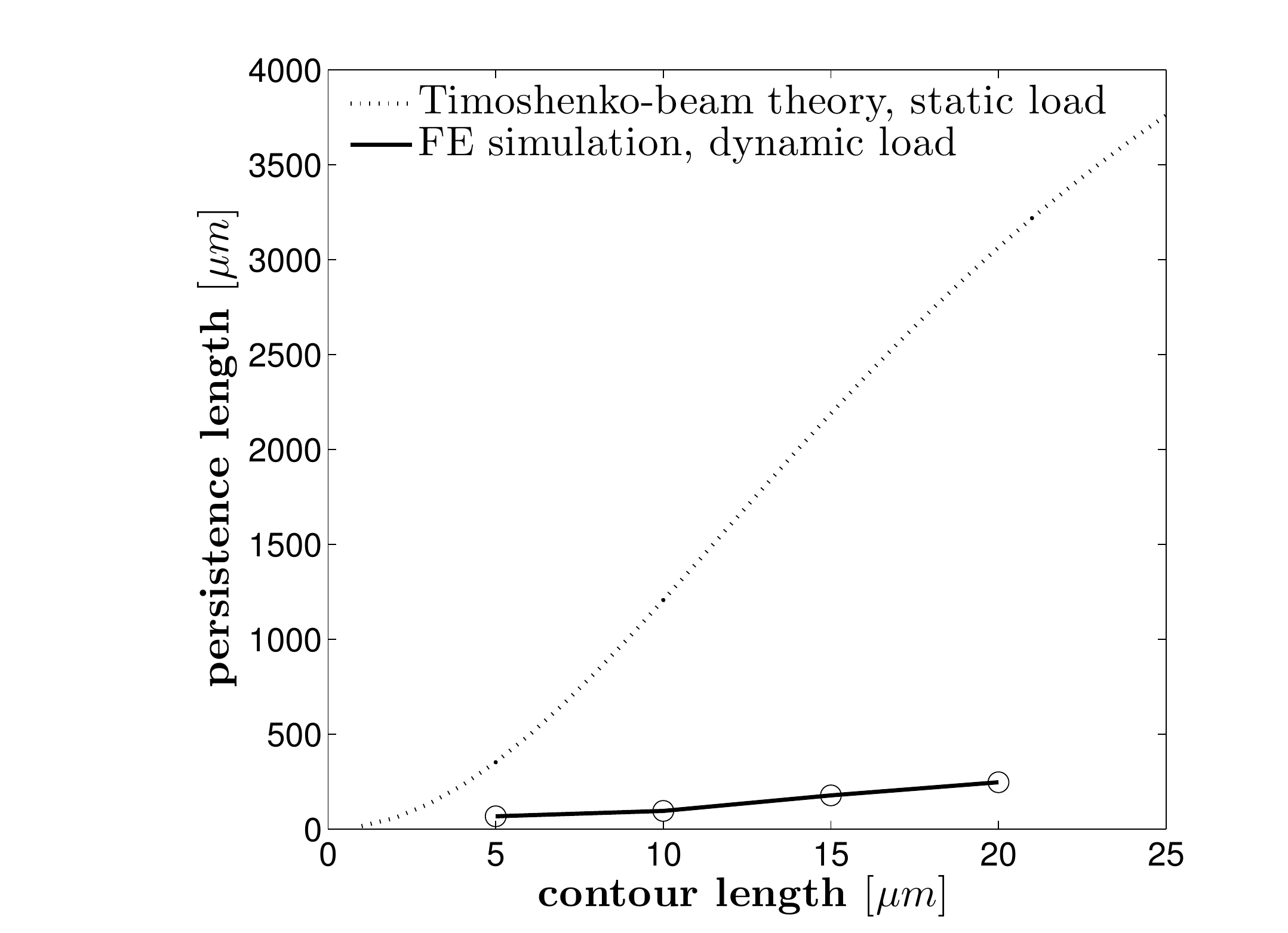}
  \caption[Comparison of simulation with static Timoshenko beam model]%
  {Persistence length versus contour length for MTs with a six-order lower longitudinal shear modulus than corresponding Young's modulus according to the static Timoshenko beam model \cite{Pampaloni2006,Taute2008} with single tip force (dashed) and dynamic FE simulations with distributed random forces (solid).}
\label{fig:chp2:comparison}
\end{figure}

We calculated the effective persistence length for MTs of various contour lengths and orders of anisotropy by means of Eq.\eqref{eq:chp2:lattanzi}. These results are presented in Figure \ref{fig:chp2:results}. We find that with increasing orders of transverse isotropy ($\chi =1,2,3$) the persistence length decreases for all contour lengths. In this range of anisotropy, short MTs seem to be less affected by a lower longitudinal shear modulus than long MTs. For highly anisotropic MTs ($\chi =6$), we see a drop in persistence length of two orders compared to their isotropic counterpart. Although changes in persistence length with contour length are seen for all values of $\chi $, this dependency is monotonic only for $\chi =6$. For this value, a gentle, but noticeable, increase of the persistence length with increasing contour length is observed. In Figure \ref{fig:chp2:comparison}, the contour length-dependency of the persistence length of MTs with the highest level of anisotropy ($\chi =6$) of the FE simulations is compared to the simplified cantilever beam model, based on the Timoshenko beam formalism and single static tip load, as used in other studies \cite{Pampaloni2006,Taute2008}.
\section{Discussion}
Certain biopolymers, such as MTs, display an anisotropic material response, which is inherent to their molecular structure  \cite{Gittes1993,Howard2001}. Unfortunately, the effective persistence length of such anisotropic filaments cannot be explicitly expressed in terms of their material properties. Additionally, the anisotropic material properties and relative thickness of MTs cannot be accurately implemented in conventional numerical and analytical models, since such models are inherently built on slenderness and isotropy approximations. Therefore, researchers have been compelled to simple static models to interpret their experimental observations. For example, based on such a static model, unexplained length-dependency of the persistence length has been related to anomalously high levels of anisotropy \cite{Pampaloni2006}. However, the validity of such static models, which are based on small angle approximations and a single deterministic tip force, may be questionable when describing the finite displacements that are typically observed for highly anisotropic filaments under distributed random forces.
To overcome these limitations we developed and validated a framework, based on the FE method set out by Cyron and Wall in \cite{Cyron2009}. This FE framework accounts for the interaction between semiflexible polymers and solvent molecules by random force generation and application. The FE model also allows for the implementation of advanced material models and nonlinear elasticity. Additionally, the developed framework is computationally efficient compared to conventionally used bead models. By applying the FE framework to isotropic MTs of various contour lengths, we found that the time evolution of thermal fluctuations is in good agreement with the predictions of the WLC model \cite{Hallatschek2004,Granek1997}. Furthermore, the FE framework allowed us to implement various levels of material anisotropy. By studying the distribution function of the tip of grafted isotropic and anisotropic MTs, we calculated the persistence length for four different contour lengths, $(L_{c}=5,10,15,20\text{ }\SI{}{\micro\meter})$. For isotropic MTs with a contour length of $L_{c}=\SI{10}{\micro\meter}$, we found a good  correspondence with theoretical predictions and other Monte Carlo studies \cite{Lattanzi2004}. 

Three trends were identified from the simulation results. First, the implementation of a transversely isotropic material model with a lower longitudinal shear modulus than corresponding Young's modulus, results in a decrease of the persistence length for all contour lengths. Second, for the first three orders of transverse isotropy ($\chi =1,2,3$), this reduction in persistence length is greater for long MTs than for short MTs. Third, implementing the highest order of transverse isotropy ($\chi =6$) yields a two order decrease in persistence length, as compared to the isotropic case. Additionally, for $\chi =6$, a slight increase of the persistence length with increasing contour length is observed.

The two orders lower persistence length of highly anisotropic MTs ($\chi=6$) is of the same order as the persistence length reported in experimental studies for short and very short MTs \cite{Pampaloni2006,VandenHeuvel2008}. These experimental results have been explained by considering MTs as weakly-coupled (at intermediate length scales \cite{Taute2008}) or decoupled (at short length scales \cite{VandenHeuvel2008}) assemblies of protofilaments. We can thus confirm that accounting for inter-protofilament decoupling by lowering the longitudinal shear modulus indeed results in such short persistence lengths.  However, the above-mentioned experiments also confirmed a significantly longer persistence length of $4-8$ \SI{}{\milli\meter} for long MTs of $L_{c}=20-40$ \SI{}{\milli\meter}, as also established in earlier studies, e.g. \cite{Gittes1993}. Such a striking increase in persistence length has been attributed to the presumption that, in this length regime, MTs behave as fully coupled homogenous structures with negligible inter-protofilament sliding. From our simulation results, we conclude that a six order difference in longitudinal shear modulus and corresponding Young's modulus alone cannot account for this reported length-dependence of the persistence length. Our findings agree well with previous research on MT rigidity, based on experimentally measurement of buckling force [33, 34] and mode decomposition of free-floating MTs \cite{Gittes1993}, in which no significant length-dependency of flexural rigidity was reported. Furthermore, the outcomes of the current study reveal the limitations of the simplified mechanical cantilever model that was used in other studies \cite{Pampaloni2006,Taute2008} to interpret the observed length-dependency. Such a static Timoshenko beam model is based on small angel approximations and a single tip load. Therefore, it is unsuitable for accurately capturing the intricate dynamics and large displacements of a highly anisotropic MTs in equilibrium with their fluid environment. The current research sheds new light on the unexplained discrepancy between the indirectly deduced longitudinal shear modulus (six order difference with corresponding Young's modulus), based on such a simplified mechanical model, and direct experimental and computational measurements of the longitudinal shear modulus \cite{Kis2002,Sept2010} (up to three orders difference with corresponding Young's modulus). The outcomes of the current study should be interpreted as an encouragement to consider other causes for the experimentally observed length-dependence than high anisotropy alone. 

The current study is subject to several limitations. A first limitation is that internal friction due to liquid flowing through narrow pores of the MT, as pointed out in \cite{Taute2008}, was neglected. Including internal friction would have caused a sharp peak in the drag coefficient for very short MTs. However, in the range of contour lengths considered in the current study, the drag coefficient may be presumed constant \cite{Taute2008}. A second limitation of the current study is that we approximated friction by an isotropic friction model. Although an isotropic friction model serves as a good first approximation, \cite{Cyron2009} an anisotropic friction model, that takes into account different friction coefficients perpendicular and longitudinal to the filament, enhances the accuracy of the absolute values of the model, see \cite{Cyron2010}. Additionally, we only implicitly accounted for the hydrodynamic interactions between distant polymer sections through scaling of the drag coefficient by a logarithmic length-dependent correction factor. Indeed, we noticed small differences in the amplitude of the saturation plateau of the MSD of the end-to-end distance upon changing the drag coefficient correction factor. Such an approximation is common practice for stiff polymers, because for such stiff polymers remote interactions, such as one segment shielding another segment, will be negligible \cite{Underhill2006, Cox1970}. However, for long and flexible polymers, hydrodynamic interaction between segments as well as self-avoidance may become appreciable and should be taken into account. In the current research the most flexible polymer ($L_{c}=\SI{20}{\micro\meter}$, $\chi=6$), was still in the stiff regime, $\frac{\ell_{p}}{L_{c}} < 0.1$. Therefore, it is expected that neglecting hydrodynamic interaction and self-avoidance will only have had a minor effect on the simulated  values and a negligible effect on the observed trends \cite{ Lattanzi2004}. Finally, due to high computational costs, the number of realizations that were performed per ensemble was limited. A higher number of realizations would have further improved the accuracy. 

Despite the above mentioned limitations, the accuracy of the model exceeds largely the required accuracy to support our main finding: a high anisotropy cannot account for the large length-dependence of the persistence length.

\bibliography{Pub_bib}

\end{document}